\documentclass[journal,12pt,letterpaper,draftcls,onecolumn]{IEEEtran}%
\usepackage{amsmath}
\usepackage{amsfonts}
\usepackage{cite}
\usepackage{amssymb}
\usepackage{graphicx}
\usepackage[usenames,dvipsnames]{color}
\usepackage{epstopdf}
\usepackage[all]{xy}
\usepackage[]{mcode}

\providecommand{\U}[1]{\protect\rule{.1in}{.1in}}
\pdfoutput=1
\providecommand{\U}[1]{\protect\rule{.1in}{.1in}}

\newtheorem{lemma}{Lemma}

\newcommand{\qed}{\nobreak \ifvmode \relax \else
      \ifdim\lastskip<1.5em \hskip-\lastskip
      \hskip1.5em plus0em minus0.5em \fi \nobreak
      \vrule height0.75em width0.5em depth0.25em\fi}

\begin{document}

\title{Outage Performance of Two-Hop OFDM Systems with Spatially Random Decode-and-Forward Relays}
\author{Shuping Dang, \IEEEmembership{Student Member, IEEE}, Justin P. Coon, \IEEEmembership{Senior Member, IEEE}, and Gaojie Chen, \IEEEmembership{Member, IEEE} 
  \thanks{
  This work was supported by the SEN grant (EPSRC grant number EP/N002350/1) and the grant from China Scholarship Council  (No. 201508060323).

   The authors are with the Department of Engineering Science, University of Oxford, Parks Road, Oxford, U.K., OX1 3PJ; tel: +44 (0)1865 283 393, (e-mail: \{shuping.dang, justin.coon, gaojie.chen\}@eng.ox.ac.uk).}}

\maketitle

\begin{abstract}
In this paper, we analyze the outage performance of different multicarrier relay selection schemes for two-hop orthogonal frequency-division multiplexing (OFDM) systems in a Poisson field of relays. In particular, special emphasis is placed on decode-and-forward (DF) relay systems, equipped with bulk and per-subcarrier selection schemes, respectively. The exact expressions for outage probability are derived in integrals for general cases. In addition, asymptotic expressions for outage probability in the high signal-to-noise ratio (SNR) region in the finite circle relay distribution region are determined in closed forms for both relay selection schemes. Also, the outage probabilities for free space in the infinite relay distribution region are derived in closed forms. Meanwhile, a series of important properties related to cooperative systems in random networks are investigated, including diversity, outage probability ratio of two selection schemes and optimization of the number of subcarriers in terms of system throughput. All analysis is numerically verified by simulations. Finally, a framework for analyzing the outage performance of OFDM systems with spatially random relays is constructed, which can be easily modified to analyze other similar cases with different forwarding protocols, location distributions and/or channel conditions.
\end{abstract}


\section{Introduction}
\IEEEPARstart{C}{ooperative} communications have become an important topic for research and industry in recent years \cite{6211487,7060199,1362898,1341264}. It is well known that relay-assisted cooperative communications are capable of providing  extra diversity and thus a better system performance in terms of energy efficiency, outage performance and network coverage extension \cite{6182883,4745938,6189409}. In particular, multicarrier relay systems are of high importance, because it fits a number of applications in practice \cite{4342860}. A number of representative and useful  multicarrier relay systems have been proposed and analyzed. For example, a block-based orthogonal frequency-division multiplexing (OFDM) decode-and-forward (DF) relay system has been analyzed in \cite{5340982}. Bulk and per-subcarrier relay selection schemes for OFDM systems have been proposed and compared in \cite{4489212} and \cite{7390520}, respectively. However, all of the above achievements regarding cooperative OFDM systems do not consider the location distribution of relays. The conventional network model employed in these previous OFDM-related works assumes the locations of all nodes to be deterministic and stationary, which form a stationary network topology. In practice, however, the dynamic nature of communication nodes is common and should be considered in order to provide a more general and meaningful analysis \cite{5226957}. Hence, a more realistic way to model a communication network is to assume the location of a node to be a random variable. To perform the analysis effectively, Poisson point processes (PPPs) have been used to analyze the location distribution of communication nodes for a large number of applications in wireless communications \cite{haenggi2013stochastic,6515339,7797167,7828097}. Pioneering work related to cooperative transmission in Poisson distributed networks was published in \cite{1705945}, in which an upper bound on outage probability is derived. Then, generalized analyses of DF and AF cooperative systems with spatially random relays distributed within a finite region have been given in \cite{6042306,5706435} and \cite{6328206,6831748}, respectively. The system with relays distributed over an infinite space is analyzed in \cite{6096785,7037462,6515497}. Opportunistic relaying with different combining techniques can be found in \cite{7248672,razi2015outage,6123785}.

However, to the best of the authors' knowledge, the link between OFDM systems and randomly distributed networks is lacking. This motivates us to construct a framework for analyzing the outage performance of two-hop OFDM systems with spatially random relays. In this paper, we analyze the outage performance of the two-hop OFDM system with spatially random DF relays and investigate a series of important properties of cooperative systems in random networks related to the outage performance. Specifically, the contributions of this paper are summarized infra:
\begin{itemize}
\item The exact expressions for outage probability for bulk and per-subcarrier selections are derived  in integral forms for general cases. Meanwhile, the asymptotic expressions for finite-region-based outage probability in the high signal-to-noise (SNR) region for bulk and per-subcarrier selections are determined in closed forms. Furthermore, the exact expressions for infinite-region-based outage probability are determined in closed forms for the case of free space.
\item It is proved that the cooperative diversity gain in Poisson random networks can either be zero, one or infinite, which is termed the \textit{ternary property}.
\item An approximate relation between the outage probability ratio of two selection schemes and the relay node density is determined, which can be used to evaluate the performance advantage of per-subcarrier selection over bulk selection in sparse networks.
\item The relation among system throughput, the number of subcarriers and relay node density is investigated and a concave problem is formulated and proved to be capable of producing the optimal number of subcarriers, so that the system throughput can be maximized. Meanwhile, a special optimization case with reliability requirement is  discussed and an approximation of the cut-off relay node density above which the formulated problem is solvable, is also derived.
\end{itemize}

All analysis is numerically verified by simulations. The results provided in this paper can be easily modified to analyze other similar cases with different forwarding protocols, location distributions and/or channel conditions.

The rest of this paper is organized as follows. The system model is detailed in Section \ref{sm}. We subsequently analyze the outage performance and discuss a series of related system properties in Section \ref{opa} and Section \ref{disp}. After that, the analysis is numerically verified by simulations in Section \ref{nr}. Finally, Section \ref{c} concludes the paper.

\section{System Model}\label{sm}
\subsection{System configurations and channel model}

In this paper, we consider a network with a single source located at the origin denoted by $\mathbf{p}_S=(0,0)$ and a destination node located at $\mathbf{p}_D=(r_{SD},0)$ in a two-dimensional polar coordinate system. The locations of source and destination are deterministic and stationary. Then, we assume the relays are homogeneously Poisson distributed over a two-dimensional region $\mathcal{C}\subseteq\mathbb{R}^2$ with a constant density $\lambda$, which form a homogeneous PPP denoted as $\Pi(\mathcal{C})$. In particular, a finite circle distribution region centered at the source node with a radius $\varsigma$ and an infinite distribution region are considered and employed to analyze the system performance in this paper, which are denoted by $\mathcal{C}_{\varsigma}$ and $\mathcal{C}_{\mathrm{inf}}$, respectively. Besides, for a typical OFDM system, we assume the number of subcarriers is $K$, which is deterministic. The set of all subcarriers is denoted as $\mathcal{K}$. Furthermore, it is assumed that the channel state information (CSI) can be perfectly estimated without any delay and overhead by the source node, so that relay selection can be effectively performed. We further suppose that the entire network operates in a half-duplex protocol and there is not a direct transmission link between source and destination due to deep fading, so that two orthogonal phases are required for one complete transmission from source to destination. In particular, the source broadcasts the signal to all relays at the first phase and relays decode and forward the received signal to the destination\footnote{We choose DF forwarding protocol in this paper due to its low CSI estimation complexity and satisfactory outage performance \cite{7504171}.}. For the noise, it is assumed to be independent and identically distributed (i.i.d.) at all nodes with noise power $N_0$.

Meanwhile, two signal degradation mechanisms encountered in transmission are considered, which are signal attenuation and multipath fading. Assuming randomly distributed relays are organized in the set $\mathcal{M}$, for the $m$th relay located at $\mathbf{p}_m=(r_{Sm},\theta_{m})$, $\forall~m\in\mathcal{M}$, if equal power allocation scheme is applied over all communication nodes with transmit power $P_t$, the received instantaneous SNR on the $k$th subcarrier is 
\begin{equation}\small
\gamma_{1}(m,k)={P_t G_{1}(m,k)r_{Sm}^{-\alpha}}/{N_0},
\end{equation}
where $\alpha$ is the path loss exponent; $G_{i}(m,k)$ is the $i$th hop channel gain on the $k$th subcarrier due to multipath fading and is modeled as an i.i.d. exponentially distributed random variable with unit mean. Therefore, for $i\in\{1,2\}$, $m\in\mathcal{M}$ and $k\in\mathcal{K}$, the probability density function (PDF) and cumulative distribution function (CDF) of $G_i(m,k)$ are given by
\begin{equation}\label{dsad4sad211}\small
f_{G}(s)=e^{-s}~\Leftrightarrow~F_{G}(s)=1-e^{-s}.
\end{equation}

For the second phase, because of the DF forwarding protocol, the received instantaneous SNR at the destination is
\begin{equation}\small
\gamma_{2}(m,k)={P_t G_{2}(m,k)r_{mD}^{-\alpha}}/{N_0},
\end{equation}
where $r_{mD}$ is the distance between the $m$th relay and destination; however, it should be noted that because we assume the locations of all relays are unchanged during a complete transmission process, $r_{mD}$ is a dependent random variable on $r_{Sm}$ and $\theta_m$, which can be expressed by the law of cosines as
\begin{equation}\label{dsakjekl12coslawwww}\small
r_{mD}=\sqrt{r_{SD}^2+r_{Sm}^2-2r_{SD}r_{Sm}\cos\theta_{m}}.
\end{equation}

Finally, the equivalent end-to-end SNR in DF relaying network can be regarded as\footnote{An outage in DF relaying networks depends on the minimum channel coefficient among the source-relay and the relay-destination links. Hence, we can employ the minimum single-hop channel SNR as the equivalent end-to-end SNR here \cite{7445895}.}
\begin{equation}\label{dsakjhe1dfdfdf}\small
\gamma(m,k)=\min\{\gamma_{1}(m,k),\gamma_{2}(m,k)\}.
\end{equation}

\begin{figure}[!t]
\centering
\includegraphics[width=4.0in]{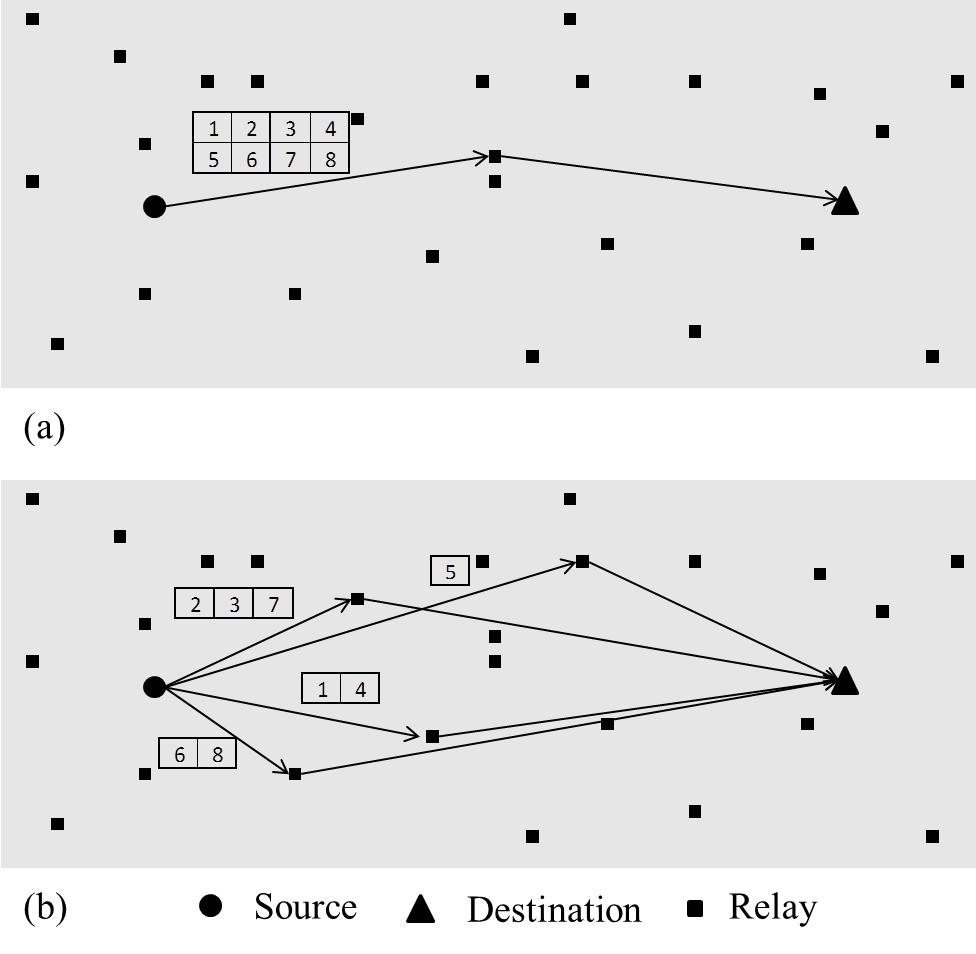}
\caption{Illustration of (a) bulk and (b) per-subcarrier relay selection schemes for single source, single destination and multiple spatially randomly distributed relays, given $K = 8$. The numbers in boxes are the sequence numbers of subcarriers.}
\label{sys}
\end{figure}

\subsection{Relay selection schemes and outage probability}
Two selection schemes are considered in the paper. First, only \textit{one} relay among $\mathcal{M}$ is selected by the selection criterion below:
\begin{equation}\label{bulkdsak2s}\small
\mathcal{L}^{bulk}=\arg\max_{m\in\mathcal{M}}\min_{k\in\mathcal{K}}\gamma(m,k).
\end{equation}

This selection scheme is termed \textit{bulk selection}, because all subcarriers will be forwarded by the only one selected relay in bulk. Obviously, this selection scheme is easy to implement for OFDM systems, since only one relay is involved in the entire transmission process. However, its outage performance is obviously not optimal for each individual subcarrier. To obtain the optimal outage performance, we can apply another selection scheme termed \textit{per-subcarrier selection}, in which multiple relays are selected in a per-subcarrier manner, so that all subcarriers can be forwarded by their optimal relays. The per-subcarrier selection criterion is given as follows\footnote{Note that, here it is allowed that $\arg\max_{m\in\mathcal{M}}\gamma(m,k)=\arg\max_{m\in\mathcal{M}}\gamma(m,n)$ for $k\neq n$. In other words, the relay is capable of forwarding two or more subcarriers simultaneously. Because we do not consider a transmit power limit in all nodes, the power imbalance problem among relays is out of the scope of this paper.}
\begin{equation}\label{psscheme1}\small
\mathcal{L}^{ps}=\bigcup_{k=1}^K \left\lbrace \arg\max_{m\in\mathcal{M}}\gamma(m,k)\right\rbrace.
\end{equation}
For clarity, these two selection schemes are illustrated in Fig. \ref{sys}.

To consider the outage performance of multiple subcarriers as an entity, let the superscript $\Xi\in\{bulk, ps\}$ standing for different relay selection schemes and define the \textit{a posteriori} outage probability after selection as\footnote{$\mathcal{M}=\varnothing$ is possible when the area of relay distribution region $|\mathcal{C}|<\infty$ and this special scenario should be regarded as outage as well \cite{6042306}.}
\begin{equation}\label{outagedefdhe89514sas}\small
\Phi^{\Xi}(s)= \mathbb{P}\left\lbrace\left\lbrace\min_{k\in\mathcal{K}} \max_{m_k\in\mathcal{L}^{\Xi}} \gamma(m_k, k)<s\right\rbrace\bigcup\left\lbrace\mathcal{M}=\varnothing\right\rbrace\right\rbrace,
\end{equation}
where $\mathbb{P}\{\cdot\}$ denotes the probability of the random event enclosed; $s$ is a predefined and fixed target SNR threshold ; $m_k$ is the index of the selected relay forwarding the $k$th subcarrier. In this paper, we will take the outage probability defined above as a metric to evaluate the outage performance.

\section{Outage Performance Analysis}\label{opa}
\subsection{Bulk selection}
By (\ref{dsad4sad211}) and (\ref{dsakjhe1dfdfdf}), the CDF of the end-to-end SNR $\gamma(m,k)$ can be derived as
\begin{equation}\label{54651243111}\small
F(s)=1-\mathrm{exp}\left(-\frac{sN_0}{P_t}\left(r_{Sm}^{\alpha}+r_{mD}^\alpha\right)\right),
\end{equation}
and by (\ref{bulkdsak2s}) the a posteriori outage probability after performing bulk selection can be determined by\footnote{The average over $\Pi(\mathcal{C})$ includes the case of $\mathcal{M}=\varnothing$ when $|\mathcal{C}|$ is finite; The area integral given in the second line of the equation can be converted to a double integral by adopting a certain coordinate system (Cartesian, polar or biangular) and thus numerically calculated.}
\begin{equation}\label{phissbulk}\small
\begin{split}
\Phi^{bulk}(s)&=\underset{\Pi(\mathcal{C})}{\mathbb{E}}\left\lbrace\prod_{m\in\mathcal{M}}\left[ 1-(1-F(s))^K\right]\right\rbrace=\mathrm{exp}\left(-\lambda\int_{\mathcal{C}}(1-F(s))^K\mathrm{d}\mathbf{p}_m\right),
\end{split}
\end{equation}
where ${\mathbb{E}}\left\lbrace\cdot\right\rbrace$ denotes the average of the enclosed.

\subsubsection{Finite relay distribution region}
Due to the symmetry of the finite circle region $\mathcal{C}_{\varsigma}$, we can derive the outage probability in the finite circle region by (\ref{phissbulk}) and obtain:
\begin{equation}\label{dsa524651423}\small
\Phi_{\varsigma}^{bulk}(s)=\mathrm{exp}\left(-2\lambda u(\varsigma)\right),
\end{equation}
where $u(\varsigma)=\int_{0}^{\pi}\int_{0}^{\varsigma}\mathcal{H}(K)\mathrm{d}r_{Sm}\mathrm{d}\theta_m$ and $\mathcal{H}(K)=r_{Sm}\mathrm{exp}\left(-\frac{KsN_0}{P_t}\left(r_{Sm}^{\alpha}+r_{mD}^{\alpha}\right)\right)$.

Although there is no closed-form expression of (\ref{dsa524651423}) because of the double integral in $u(\varsigma)$, we can employ a power series expansion at $\frac{P_t}{N_0}\rightarrow \infty$ and obtain the asymptotic expressions of $\Phi^{bulk}_{\varsigma}(s)$ for an arbitrary $\alpha$:

\begin{equation}\label{jianjinhaoc}\small
\Phi_{\varsigma}^{bulk}(s)\sim\tilde\Phi_{\varsigma}^{bulk}(s)=\mathrm{exp}\left(-\lambda\pi\varsigma^2\left(1-\frac{KsN_0\tau_\alpha}{P_t}\right)\right),
\end{equation}
where 
\begin{equation}\small
\tau_{\alpha}=
\begin{cases}
r_{SD}^2+\varsigma^2,~~~~~~~~~~~~~~~~~~~~~~~~~~~~~~~~~~~~~\alpha=2\\
r_{SD}^4+2r_{SD}^2\varsigma^2+\frac{2}{3}\varsigma^4,~~~~~~~~~~~~~~~~~~~~~~~\alpha=4\\
\frac{1}{2}\left(2r_{SD}^2+\varsigma^2\right)\left(r_{SD}^4+4r_{SD}^2\varsigma^2+\varsigma^4\right),~~~~\alpha=6
\end{cases}.
\end{equation}
From (\ref{jianjinhaoc}), it can be seen that with an increasing $\frac{P_t}{N_0}$, $\Phi_{\varsigma}^{bulk}(s)$ will converge to the outage floor given by
\begin{equation}\small
\underline \Phi_{\varsigma}^{bulk}(s)=\mathrm{exp}(-\lambda\pi\varsigma^2),
\end{equation}
which is caused by the scenario where there is no relay existing in the finite circle distribution region (c.f. (\ref{outagedefdhe89514sas})).

\subsubsection{Infinite relay distribution region}
Due to the symmetry of the infinite region $\mathcal{C}_{\mathrm{inf}}$, we can  obtain
\begin{equation}\label{haochangya1}\small
\begin{split}
\Phi_{\mathrm{inf}}^{bulk}(s)=\mathrm{exp}\left(-2\lambda\int_{0}^{\pi}\int_{0}^{\infty}\mathcal{H}(K)\mathrm{d}r_{Sm}\mathrm{d}\theta_m\right)
\end{split}.
\end{equation}
To the best of the authors' knowledge, there does not exist a closed form of (\ref{haochangya1}) for an arbitrary $\alpha$. However, for a special case when $\alpha=2$ (free space), we can obtain
\begin{equation}\label{haochangya2}\small
\Phi_{\mathrm{inf}}^{bulk}(s)\vert_{\alpha=2}=\mathrm{exp}\left(-\frac{\lambda\pi P_t}{2KsN_0}\mathrm{exp}\left(-\frac{r_{SD}^2KsN_0}{2P_t}\right)\right).
\end{equation}

\subsection{Per-subcarrier selection}
Similarly as the case of bulk selection, by (\ref{psscheme1}) and the binomial theorem, the a posteriori outage probability after performing per-subcarrier selection can be determined by
\begin{equation}\label{phissps}\small
\begin{split}
&\Phi^{ps}(s)=\underset{\Pi(\mathcal{C})}{\mathbb{E}}\left\lbrace1-\left(1-\prod_{m\in\mathcal{M}}F(s)\right)^K\right\rbrace\\
&=\sum_{k=1}^{K}\left[\binom{K}{k}(-1)^{k+1}\underset{\Pi(\mathcal{C})}{\mathbb{E}}\left\lbrace\prod_{m\in\mathcal{M}}F^k(s)\right\rbrace\right]=\sum_{k=1}^{K}\left[\binom{K}{k}(-1)^{k+1}\mathrm{exp}\left(-\lambda\int_{\mathcal{C}}\left(1-F^k(s)\right)\mathrm{d}\mathbf{p}_m\right)\right].
\end{split}
\end{equation}

\subsubsection{Finite relay distribution region}
We can obtain the outage probability of per-subcarrier selection scheme over finite relay distribution region by
\begin{equation}\label{54d65sa465d1}\small
\begin{split}
\Phi_{\varsigma}^{ps}(s)=\sum_{k=1}^{K}\left[\binom{K}{k}(-1)^{k+1} \mathrm{exp}\left(-2\lambda\sum_{n=1}^{k}\binom{k}{n} (-1)^{n+1}  u(\varsigma,n)\right)\right],
\end{split}
\end{equation}
where
$u(\varsigma,n)=\int_{0}^{\pi}\int_{0}^{\varsigma} \mathcal{H}(n)\mathrm{d}r_{Sm}\mathrm{d}\theta_{m}$.

Meanwhile, by power series expansion on (\ref{54d65sa465d1}) at $\frac{P_t}{N_0}\rightarrow \infty$, we also obtain the asymptotic expression of $\Phi_{\varsigma}^{ps}(s)$ as
\begin{equation}\label{jianjinhaocpsssss}\small
\begin{split}
&\Phi_{\varsigma}^{ps}(s)\sim\tilde\Phi_{\varsigma}^{ps}(s)=
\mathrm{exp}\left(-\lambda\pi\varsigma^2\right)\left[1-K\left(1-\mathrm{exp}\left(\frac{\lambda\pi\varsigma^2sN_0\tau_\alpha}{P_t}\right)\right)\right]
\end{split},
\end{equation}
from which we can observe that the outage floor given by
\begin{equation}\small
\begin{split}
\underline\Phi_{\varsigma}^{ps}(s)=\mathrm{exp}(-\lambda\pi\varsigma^2)=\underline\Phi_{\varsigma}^{bulk}(s),
\end{split}
\end{equation}
is exactly the same as the case of bulk selection. As expected, the outage event at high SNR within a finite circle distribution region is dominated by the case where there is no relay candidate for selection (i.e. $\mathcal{M}=\varnothing$). As a result, it is irrelevant to the selection schemes and/or forwarding protocols and only dependent on the relay node density $\lambda$ and the radius $\varsigma$. In general, the outage floor for an arbitrary finite distribution region $\mathcal{C}$ can be derived by
\begin{equation}\small
\underline\Phi(\mathcal{C})=\mathrm{exp}\left(-\lambda|\mathcal{C}|\right).
\end{equation}

\subsubsection{Infinite relay distribution region}
Similarly, substituting $\mathcal{C}=\mathcal{C}_{\mathrm{inf}}$ into (\ref{phissps}) yields 
\begin{equation}\label{haochangyapsdf}\small
\begin{split}
&\Phi_{\mathrm{inf}}^{ps}(s)=\sum_{k=1}^{K}\left[\binom{K}{k}(-1)^{k+1} \mathrm{exp}\left(-2\lambda\sum_{n=1}^{k}\binom{k}{n} (-1)^{n+1}  \int_{0}^{\pi}\int_{0}^{\infty} \mathcal{H}(n) \mathrm{d}r_{Sm}\mathrm{d}\theta_{m}\right)\right].
\end{split}
\end{equation}

Again, there is no closed form of (\ref{haochangyapsdf}) for an arbitrary $\alpha$. When $\alpha=2$, we can obtain
\begin{equation}\label{218sjsyrtysy3}\small
\begin{split}
&\Phi_{\mathrm{inf}}^{ps}(s)\vert_{\alpha=2}=\sum_{k=1}^{K}\left[\binom{K}{k}(-1)^{k+1} \mathrm{exp}\left(-\lambda\pi\sum_{n=1}^{k}\binom{k}{n}   (-1)^{n+1}\frac{ P_t}{2 nsN_0}\mathrm{exp}\left(-\frac{r_{SD}^2nsN_0}{2P_t}\right) \right)\right].
\end{split}
\end{equation}

\section{Discussion of Important System Properties}\label{disp}
\subsection{Cooperative diversity analysis}
For deterministic networks in which the number of relays and/or their locations are stationary, diversity gain is an all-important metric to measure the cooperative performance advantage \cite{1362898}. However, we show as follows that this metric is not appropriate anymore for cooperative systems in random networks where relays are Poisson distributed. Subsequently we prove the \textit{ternary property} of diversity gain in Poisson random networks.

For cooperative systems in finite Poisson random networks, the binary property of diversity gain has been proved and it states that the diversity gain $d_o(\mathcal{C})$ can only be either one or zero depending on whether there is a direct transmission link between source and destination or not \cite{6042306}. Following the discussion of outage floor in Section \ref{opa}, this binary property related to diversity gain is straightforward. Because without a direct transmission link, an increasing power would not affect the outage performance at high SNR and this leads to a \textit{zero-diversity-order system}.

However, when considering an infinite distribution region, there does not exist such an outage floor, because if the region is infinite, as long as the relay node density $\lambda$ is positive, there are always an infinite number of relays distributed over region $\mathcal{C}_{\mathrm{inf}}$. As a result, with a \textit{sufficiently} large transmit power (considering the asymptotic region), the two-hop transmission can always be successful, because there must exist a `satisfactory' two-hop link among an infinite number of relays. This indicates the diversity gain of cooperative systems in infinite random networks is infinity. Mathematically, for an arbitrary and bounded $\alpha$, we can prove that\footnote{See Appendix \ref{eppenproxgaindiv}.}
\begin{equation}\label{dsakj12837223gmis}\small
d_o^{\Xi}(\mathcal{C}_{\inf})=-\lim_{\frac{P_t}{N_0}\rightarrow\infty}\frac{\log (\Phi_{\mathrm{inf}}^{\Xi}(s))}{\log (P_t/N_0)}=\infty.
\end{equation}

Therefore, unlike the situations in deterministic networks, it would be impossible to derive a linear relation among $d_o(\mathcal{C})$, $\lambda$ and $\alpha$ in random networks. Instead, the diversity gain can only be either zero, one or infinity. We term this \textit{ternary property} of diversity gain in Poisson random networks. In the meantime, (\ref{dsakj12837223gmis}) also indicates that any high but bounded path loss attenuation does not counter the constructive effects of infinite distribution region on asymptotic outage performance.

\subsection{Comparison of outage performances between bulk and per-subcarrier selections}\label{sa65d6142321312}
By (\ref{phissbulk}) and (\ref{phissps}), we can quantify the outage performance advantage of per-subcarrier selection over bulk selection via the outage probability ratio given by
\begin{equation}\label{45philamd65dsa451}\small
\phi(\lambda)=\frac{\Phi^{ps}(s)}{\Phi^{bulk}(s)}=\sum_{k=1}^{K}\left[\binom{K}{k}(-1)^{k+1}\mathrm{exp}\left(-\lambda\Delta(k)\right)\right],
\end{equation}
where $\Delta(k)=\int_{\mathcal{C}}\left[1-F^k(s)-\left(1-F(s)\right)^K\right]\mathrm{d}\mathbf{p}_m$.

Considering the case of sparse networks with small $\lambda$, the outage probability ratio $\phi(\lambda)$ can be approximated by\footnote{See Appendix \ref{5464218penglua}.}
\begin{equation}\label{jinsidephilamd}\small
\phi(\lambda)\approx 1+\frac{\lambda^2}{2}\sum_{k=1}^{K}\left[\binom{K}{k}(-1)^{k+1}\Delta^2(k)\right].
\end{equation}

Eq. (\ref{jinsidephilamd}) indicates that the outage performance advantage brought by per-subcarrier selection over bulk selection will become negligible with a decreasing $\lambda$, because the number of relays for selection is small and employing multiple relays to perform per-subcarrier selection becomes less likely. Meanwhile, if we would like to ensure the performance advantage by letting the outage probability ratio $\phi(\lambda)\leq\epsilon$ (i.e. the outage probability of per-subcarrrier selection is $\epsilon$ times lower than that of bulk selection), an approximation of the required minimum density $\lambda$ can be derived as
\begin{equation}\label{solutlamndsa2}\small
\lambda\geq\underline{\lambda}(\epsilon)\approx\sqrt{{2(\epsilon-1)}/{\sum_{k=1}^{K}\left[\binom{K}{k}(-1)^{k+1}\Delta^2(k)\right]}},
\end{equation}
which would provide a guidance for whether bulk or per-subcarrier selection should be employed or a switching criterion for a dynamic switching mechanism between these two selection schemes when the relay node density $\lambda$ is known\footnote{Although per-subcarrier selection scheme has the optimal outage performance, it is not always preferable if the performance advantage is not significant, because the selection and synchronization processes among multiple relays and destination node are complicated and thus will require more overheads \cite{6356932}.}.

\subsection{Optimization of the number of subcarriers for bulk selection}
With the development of index modulation and frequency resource allocation, OFDM systems with a varying number of active subcarriers become common in practice \cite{6587554,7809043}. Therefore, it is meaningful to investigate the effects of adopting different numbers of subcarriers $K$ on the system throughput. For simplicity, we assume that equal numbers of bits are carried by each subcarrier. Hence, the average system throughput can be characterized by the average number of successfully decoded subcarriers per transmission at the destination \cite{1343893}, which can be determined by
\begin{equation}\label{213112kappa}\small
\kappa(K,\lambda)=K\left(1-\Phi^{\Xi}(s)\right).
\end{equation}

\begin{figure}[!t]
\centering
\includegraphics[width=5.5in]{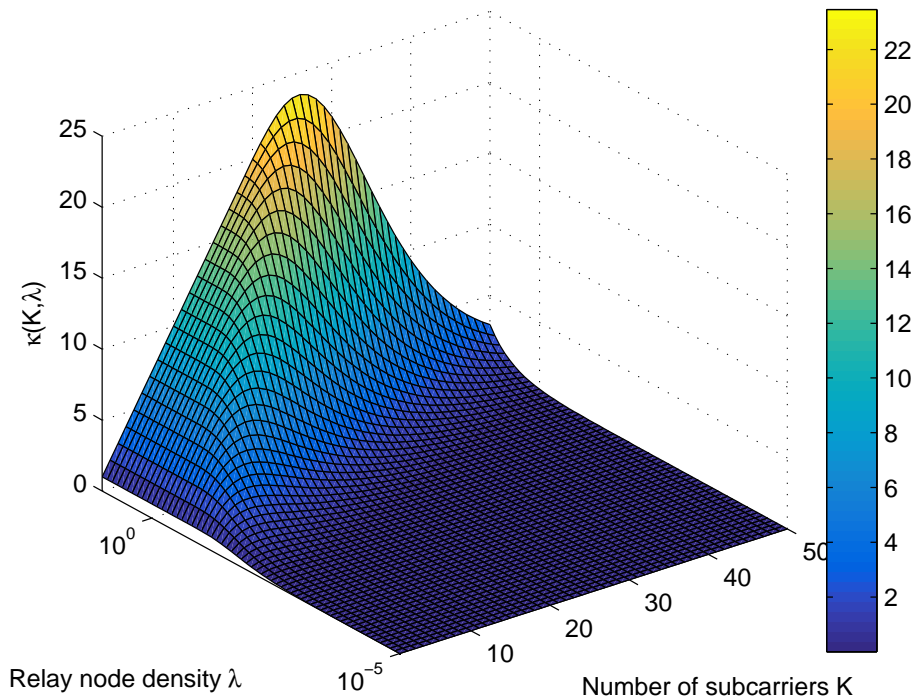}
\caption{Relation among $\kappa(K,\lambda)$, $K$ and $\lambda$ for bulk selection in free space ($\alpha=2$), when $s=1$, $P_t/N_0=100$, $r_{SD}=5$ and $\varsigma=5$.}
\label{fujia21}
\end{figure}

According to (\ref{outagedefdhe89514sas}), a large $K$ will lead to a large $\Phi^{\Xi}(s)$, since all signals transmitted on these $K$ subcarriers need to be successfully decoded at the destination, or an outage will occur otherwise. Consequently, for a given $\lambda$, there exists an optimal number of subcarriers denoted by $K_{\mathrm{opt}}$, by which the system throughput can be maximized. Also, because relay node density $\lambda$ is varying in practice due to the dynamic on/off switching mechanisms and node mobility \cite{6502479,565669}, it is intuitive that we can dynamically adjust the number of subcarriers $K$ to offset the adverse  effects of relay node density $\lambda$, so that the system throughput can always be maximized. To be clear, the relation among $\kappa(K,\lambda)$, $K$ and $\lambda$ is plotted in Fig. \ref{fujia21} for the case of bulk selection in free space. 

Now, let us focus on the method to determine $K_{\mathrm{opt}}$. First, we relax the integer $K$ to a positive real number $\tilde{K}$\footnote{Currently, this method is only applicable for bulk selection, since $K$ can be regarded as a product factor in $\Phi^{bulk}(s)$ and thus can be relaxed to $\tilde{K}$. On the other hand, because $K$ is the upper limit of the summation in $\Phi^{ps}(s)$, this relaxing relation is not feasible, and thus the following optimization is not possible for per-subcarrier selection. However, it is possible to propose a sub-optimal solution for per-subcarrier selection in which the $\Phi^{ps}(s)$ is replaced by a certain approximation with $K$ as a product factor.}, so that some optimization tools can be applied. Then, it can be proved that $\kappa(\tilde K,\lambda)$ is a concave function of $\tilde K$\footnote{See Appendix \ref{2311profconcav}}. Therefore, the optimal solution to $\tilde K$ can be obtained via a concave problem formulated by
\begin{equation}\label{dask2caveskaps}\small
\begin{split}
&\tilde K_{\mathrm{opt}}=\arg\max_{\tilde K} \kappa(\tilde K,\lambda)\\
&~~~~~~~~\mathrm{s.t.}~~\tilde K>0,
\end{split}
\end{equation}
which can be solved efficiently using standard optimization techniques (e.g. CVX in MATLAB). Then, $K_{\mathrm{opt}}$ is determined by
\begin{equation}\small
K_{\mathrm{opt}}=\begin{cases}
\lceil\tilde K_{\mathrm{opt}}\rceil,~~\mathrm{if}~\kappa(\lceil\tilde K_{\mathrm{opt}}\rceil,\lambda)\geq \kappa(\lfloor\tilde K_{\mathrm{opt}}\rfloor,\lambda)\\
\lfloor\tilde K_{\mathrm{opt}}\rfloor,~~\mathrm{if}~\kappa(\lceil\tilde K_{\mathrm{opt}}\rceil,\lambda)< \kappa(\lfloor\tilde K_{\mathrm{opt}}\rfloor,\lambda),
\end{cases}
\end{equation}
where $\lceil\cdot\rceil$ and $\lfloor\cdot\rfloor$ denote ceiling and floor functions, respectively.

However, the optimization of $\kappa(K,\lambda)$ given above would not be always applicable, since the chosen $K_{\mathrm{opt}}$ would lead to an inappropriate outage probability beyond a threshold $\Psi$, which should be maintained for a prescribed quality of service (QoS). Unlike networks for real-time streaming media in which throughput is the key \cite{7362038}, this reliability requirement is in particular crucial for some special networks, e.g. the Internet of things (IoT) \cite{atzori2010internet} and military wireless sensor networks (WSNs) \cite{5379900}. In this context, $K_{\mathrm{opt}}$ is dependent on $\lambda$ and $\Psi$. Therefore, with this constraint on outage probability $\Psi$, the concave optimization problem formulated in (\ref{dask2caveskaps}) becomes
\begin{equation}\label{dask2caves232322kaps}\small
\begin{split}
&\tilde K_{\mathrm{opt}}=\arg\max_{\tilde K} \kappa(\tilde K,\lambda)\\
&~~~~~~~~\mathrm{s.t.}~~\tilde K>0~~\mathrm{and}~~\Phi^{\Xi}(s)\vert_{K=\tilde K}\leq\Psi.
\end{split}
\end{equation}
Then $K_{\mathrm{opt}}$ will be determined by
\begin{equation}\small
K_{\mathrm{opt}}=\lfloor\tilde K_{\mathrm{opt}}\rfloor.
\end{equation}

On the other hand, because $K\geq 1$ must be ensured in order to provide transmission service, it is impossible to always maintain a given outage probability $\Psi$ by reducing $K$ when $\lambda$ keeps decreasing. Consequently, there is a cut-off relay node density $\lambda_c$ below which the outage probability $\Psi$ cannot be maintained by reducing $K$. By substituting $K=1$ into (\ref{phissbulk}), the cut-off density can be numerically evaluated by
\begin{equation}\small
\lambda_c=-\frac{\ln \Psi}{\int_{\mathcal{C}}(1-F(s))\mathrm{d}\mathbf{p}_m}.
\end{equation}
By (\ref{haochangya2}), the cut-off relay node density for bulk selection in free space can be approximated by
\begin{equation}\small
\lambda_c\vert_{\alpha=2}\approx-\frac{2sN_0\ln \Psi}{\pi P_t \mathrm{exp}\left(-\frac{r_{SD}^2sN_0}{2P_t}\right)}.
\end{equation}

\begin{figure}[!t]
\centering
\includegraphics[width=5.0in]{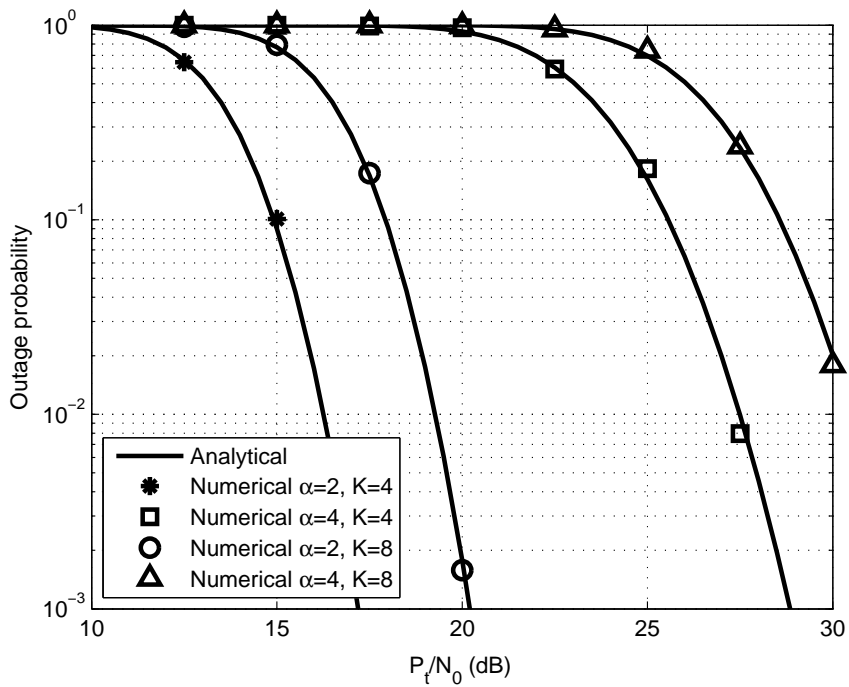}
\caption{Bulk selection case: outage probability vs. $P_t/N_0$, when $\lambda=1$, $\varsigma=5$ and $r_{SD}=5$.}
\label{tupian1}
\end{figure}

\begin{figure}[!t]
\centering
\includegraphics[width=5.0in]{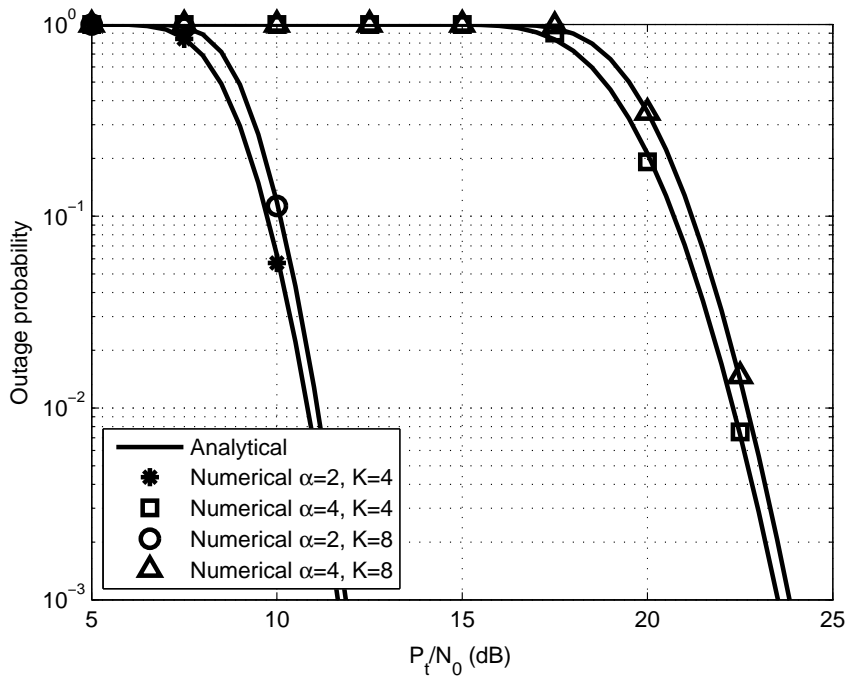}
\caption{Per-subcarrier selection case: outage probability vs. $P_t/N_0$, when $\lambda=1$, $\varsigma=5$ and $r_{SD}=5$.}
\label{tupian2}
\end{figure}

\begin{figure}[!t]
\centering
\includegraphics[width=5.0in]{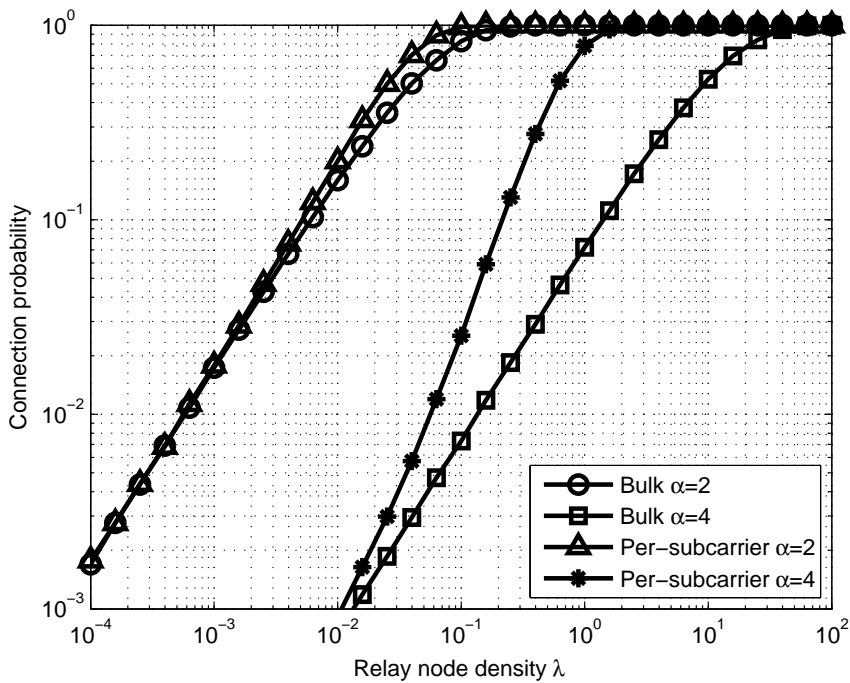}
\caption{Connection probability vs. relay node density $\lambda$, when $P_t/N_0=100$, $\varsigma=5$, $r_{SD}=5$ and $K=4$.}
\label{tupian3}
\end{figure}

\begin{figure}[!t]
\centering
\includegraphics[width=5.0in]{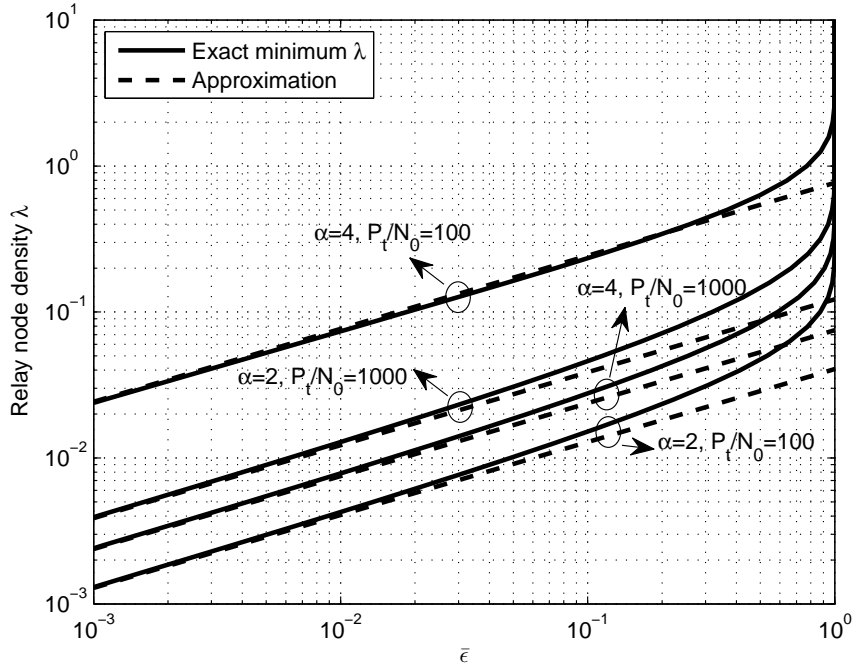}
\caption{$\bar\epsilon$ vs. relay node density $\lambda$, when $\varsigma=5$, $r_{SD}=5$ and $K=4$.}
\label{fujia1}
\end{figure}

\begin{figure}[!t]
\centering
\includegraphics[width=5.0in]{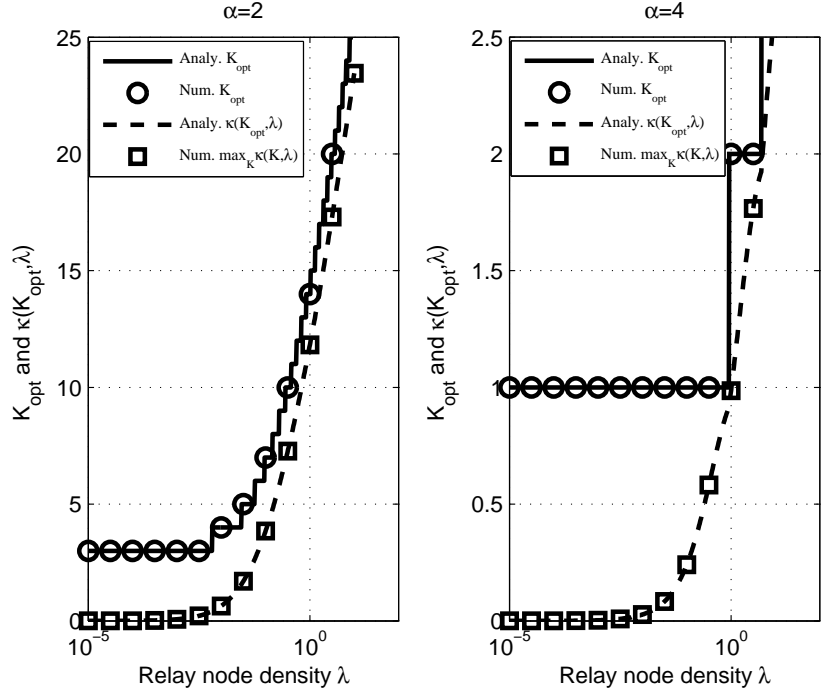}
\caption{Relation between relay node density $\lambda$ and the optimal number of subcarriers $K_{\mathrm{opt}}$ as well as the maximum system throughput characterized by $\kappa(K_{\mathrm{opt}},\lambda)$, when $s=1$, $P_t/N_0=100$, $\varsigma=5$ and $r_{SD}=5$.}
\label{kappa}
\end{figure}

\begin{figure}[!t]
\centering
\includegraphics[width=5.0in]{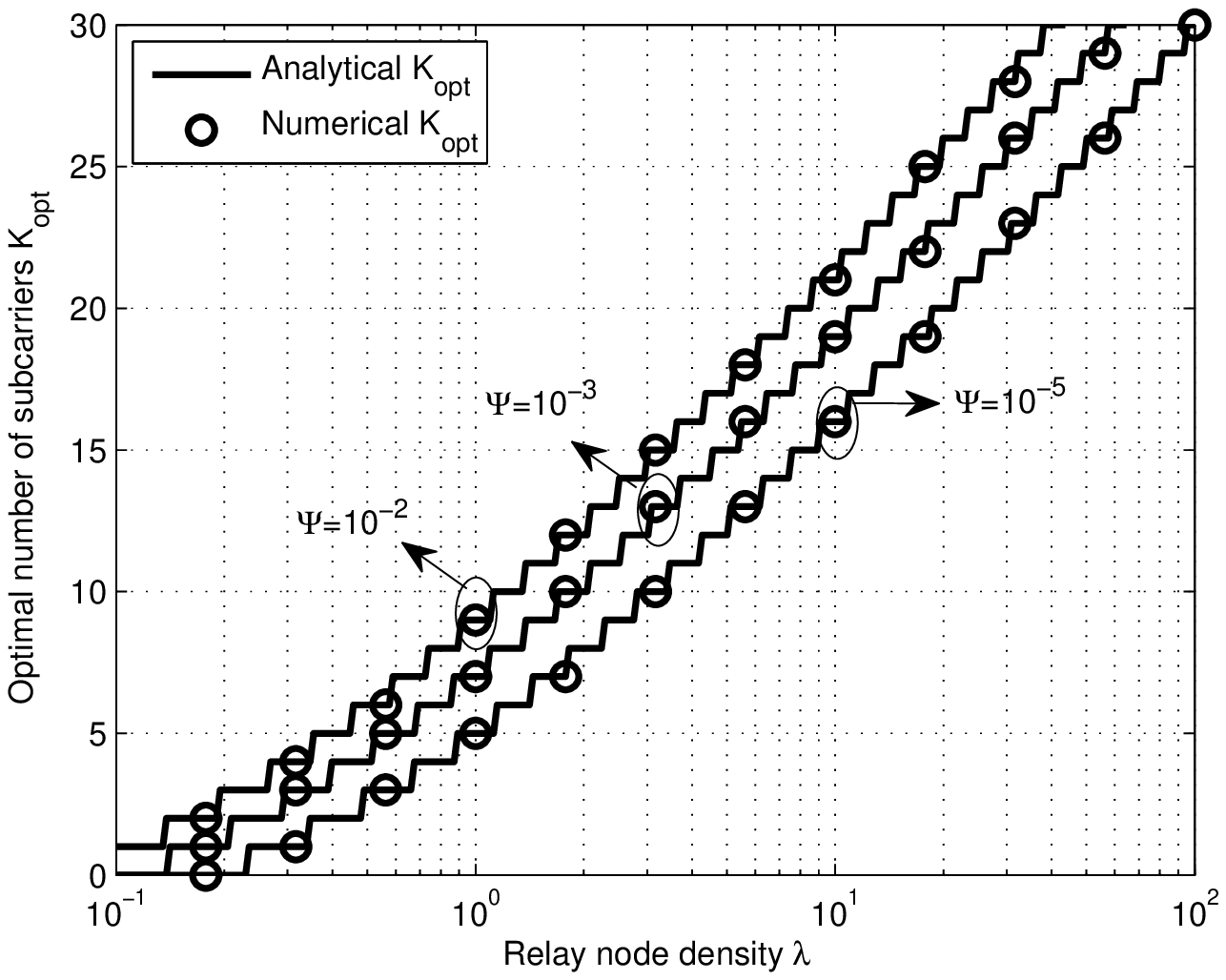}
\caption{Relation between relay node density $\lambda$ and the optimal number of subcarriers $K_{\mathrm{opt}}$ for a given outage probability $\Psi\in\{10^{-2},10^{-3},10^{-5}\}$, when $\alpha=2$, $s=1$, $P_t/N_0=100$, $\varsigma=5$ and $r_{SD}=5$.}
\label{fujia22}
\end{figure}

\section{Numerical Results}\label{nr}
\subsection{Outage performance of bulk and per-subcarrier selections}
First of all, we need to verify the correctness of the analytical results of outage probabilities. To do so, we normalize $s=1$, $N_0=1$ and set $\lambda=1$, $\varsigma=5$ and $r_{SD}=5$\footnote{The lengths  given here are relative and thus dimensionless, since other parameters have been normalized.}. Then, we simulate the relation between the outage probability and the transmit power with spatially random relays distributed over a finite circle region. The simulation results for bulk and per-subcarrier selections are presented in Fig. \ref{tupian1} and Fig. \ref{tupian2}, respectively. By the results shown in these two figures, our analysis of the outage performance corresponding to finite circle region has been verified. Meanwhile, we can also see that the variation of the number of subcarriers $K$ has a significant impact on the outage performance for bulk selection, while it is relatively trivial for per-subcarrier selection. This is because bulk selection can only employ one relay to satisfy transmission requirements of all subcarriers, which is more `$K$-sensitive' than the case of per-subcarrier selection which is capable of employing multiple relays. Both selection schemes are affected by path loss exponent in a large but similar scale, which aligns with our expectation.

\subsection{Relation between relay node density and multi-relay performance advantage}
In order to investigate the effects of the relay node density $\lambda$ on outage probability ratio $\phi(\lambda)$, we define the \textit{connection probability} as $\bar \Phi^{\Xi}(s)= 1- \Phi^{\Xi}(s)$, which eases the illustration of some crucial properties of spatially random networks in logarithmic plots when $\lambda$ is small\footnote{Because when $\lambda$ is small, the outage probability will approach to one and the logarithmic scale is a nonlinear scale in terms of the power of ten, which cannot show the details around one.}. Then, we set $P_t/N_0=100$ and $K=4$ and plot the relation between connection probability and $\lambda$ in Fig. \ref{tupian3}. From  Fig. \ref{tupian3}, it is obvious that when $\lambda$ is small (i.e. in sparse networks), the multi-relay performance advantage brought by per-subcarrier selection over bulk selection will become negligible, because the number of relays for selection is small and employing multiple relays to perform per-subcarrier selection becomes less likely. Also, in order to show the properties for small $\lambda$ in logarithmic plots, we specify $\bar\epsilon=1-\epsilon$ and plot the relation between $\bar\epsilon$ and $\lambda$ to verify the proposed approximation (\ref{solutlamndsa2}) in Fig. \ref{fujia1}\footnote{As the same reason that we define connection probability, when $\lambda$ is small, $\epsilon$ will approach to one and the details cannot be shown by the logarithmic scale based on the power of ten. In Fig. \ref{fujia1}, the exact $\lambda$ is produced by numerically inversely solving (\ref{45philamd65dsa451}) for $\phi(\lambda)=1-\bar\epsilon$, while the approximate $\lambda$ is produced by calculating $\underline{\lambda}(1-\bar{\epsilon})$, as given in (\ref{solutlamndsa2}).}. From this figure, it is verified that the proposed approximation on $\underline{\lambda}(\epsilon)$ is tight for small $\lambda$ and can thereby provide an efficient metric for evaluating the performance advantage brought by per-subcarrier selection over bulk selection in sparse networks.

\subsection{Relation between the number of subcarriers and system throughput}
To verity the solutions to the optimal number of subcarriers produced by the concave problem formulated in (\ref{dask2caveskaps}), we set $r_{SD}=5$, $\varsigma=5$ and $P_t/N_0=100$ and plot the simulation results in Fig. \ref{kappa}. The jagged nature of $K_{\mathrm{opt}}$ reflects $K$, as the number of subcarriers can only take discrete values in practice. Meanwhile, from this figure, the solutions produced by the concave problem exactly match the numerical results, which verifies the accuracy of the proposed solution. It also aligns with our expectation that with an increasing $\lambda$, both $K_{\mathrm{opt}}$ and the maximum $\kappa(K,\lambda)$ will increase, since the signal propagation environment has been improved with more `appropriate' relays. Also, a higher path loss exponent will significantly degrade the system throughput, which can be observed by comparing the results shown for $\alpha=2$ and $\alpha=4$. The degradation due to high path loss exponent is obvious especially when $\lambda$ is large.

Besides, to verify the constrained solution to the optimal number of subcarriers $K_{\mathrm{opt}}$ for a given outage probability $\Psi$, the simulation results are presented in Fig. \ref{fujia22}. Here, we specify $K_{\mathrm{opt}}=0$ when $\lambda$ is lower than the cut-off density $\lambda_c$ for simplicity and frequency resource saving purposes. Again, both numerical and analytical results match each other and these results verify the feasibility of the proposed solution for the scenario with an outage constraint. Also, in comparison with the results illustrated in Fig. \ref{kappa}, the constraint on outage probability will result in a lower optimal number of subcarriers. Furthermore, by comparing the cases with different $\Psi$, it can be found that the optimal number of subcarriers reduces with the constraint on outage.

\section{Conclusion}\label{c}
In this paper, we analyzed the outage performance of two-hop OFDM systems adopting bulk and per-subcarrier selection schemes, respectively. Also, a series of important properties related to cooperative systems in random networks were investigated, including diversity, outage probability ratio of two selection schemes and the optimization of the number of subcarriers. All analysis has been verified by simulations and some key properties of cooperative OFDM systems over finite and infinite random networks have been revealed and discussed. Moreover, by (\ref{phissbulk}) and (\ref{phissps}), an analytical framework for OFDM systems over random networks has been constructed in this paper, which can be easily modified to analyze other similar cases with different forwarding protocols, location distributions and/or channel conditions.

\appendices

\section{Proof of Diversity Gain in Infinite Poisson Random Networks}\label{eppenproxgaindiv}
To determine the diversity gain in infinite Poisson random networks for bulk selection, we first have the relation
\begin{equation}\label{dachangesdas2}\small
\begin{split}
&d_o^{bulk}(\mathcal{C}_{\mathrm{inf}})=-\lim_{\frac{P_t}{N_0}\rightarrow\infty}\frac{\log\left(\Phi_{\mathrm{inf}}^{bulk}(s)\right)}{\log\left(P_t/N_0\right)}\\
&\overset{(\mathrm{a})}{>}\lim_{\frac{P_t}{N_0}\rightarrow\infty} \frac{2\pi\lambda\int_{0}^{\infty}r_{Sm}\mathrm{exp}\left(-\frac{KsN_0}{P_t}\left(r_{Sm}^\alpha+(r_{Sm}+r_{SD})^\alpha\right)\right)\mathrm{d}r_{Sm}}{\ln 10 \log(P_t/N_0)}\\
&>\mathcal{T}_2=\lim_{\frac{P_t}{N_0}\rightarrow\infty} \frac{2\pi\lambda\mathcal{T}_1}{\ln 10 \log(P_t/N_0)},
\end{split}
\end{equation}
where $(\mathrm{a})$ holds because of the trigonometric relation among $r_{mD}$, $r_{Sm}$ and $\theta_m$ expressed in (\ref{dsakjekl12coslawwww}); $\mathcal{T}_1$ is given by
\begin{equation}\label{1123changdes}\small
\begin{split}
\mathcal{T}_1&=\int_{0}^{\infty}r_{Sm}\mathrm{exp}\left(-\frac{KsN_0}{P_t}\left(r_{Sm}^\alpha+(2\max(r_{Sm},r_{SD}))^\alpha\right)\right)\mathrm{d}r_{Sm}\\
&=\int_{0}^{r_{SD}}r_{Sm}\mathrm{exp}\left(-\frac{KsN_0}{P_t}\left(r_{Sm}^\alpha+(2r_{SD})^\alpha\right)\right)\mathrm{d}r_{Sm}+\int_{r_{SD}}^{\infty}r_{Sm}\mathrm{exp}\left(-\frac{KsN_0}{P_t}\left(r_{Sm}^\alpha+(2r_{Sm})^\alpha\right)\right)\mathrm{d}r_{Sm}\\
&=\left(\frac{P_t}{KsN_0}\right)^{\frac{2}{\alpha}}\mathrm{exp}\left(-\frac{KsN_0(2r_{SD})^\alpha}{P_t}\right)\underline{\Gamma}\left(\frac{2}{\alpha},\frac{KsN_0r_{SD}^\alpha}{P_t}\right)+\frac{r_{SD}^2}{\alpha}{E}_{\frac{\alpha-2}{\alpha}}\left(\frac{(1+2^\alpha)KsN_0r_{SD}^\alpha}{P_t}\right),
\end{split}
\end{equation}
where $\underline{\Gamma}(n,x)=\int_{0}^{x}t^{n-1}e^{-t}\mathrm{d}t$ is the lower incomplete gamma function and ${E}_{n}(x)=\int_{1}^{\infty}{e^{xt}}/{t^n}\mathrm{d}t$ is the exponential integral function.

Substituting (\ref{1123changdes}) into the last line of (\ref{dachangesdas2}) yields
\begin{equation}\small
\mathcal{T}_2=\lim_{\frac{P_t}{N_0}\rightarrow\infty} \frac{2\pi\lambda\mathcal{T}_1}{\ln 10 \log(P_t/N_0)}=\infty.
\end{equation}

As a consequence of $d_o^{bulk}(\mathcal{C}_{\mathrm{inf}})>\mathcal{T}_2$, we have $ d_o^{bulk}(\mathcal{C}_{\mathrm{inf}})=\infty$. Meanwhile, as different selection schemes will not affect diversity gain, which is only related to relay's distribution \cite{4489212}, we can also have $d_o^{ps}(\mathcal{C}_{\mathrm{inf}})=d_o^{bulk}(\mathcal{C}_{\mathrm{inf}})=\infty$. Therefore, considering an infinite distribution region $\mathcal{C}_{\mathrm{inf}}$, for $\Xi\in\{bulk,ps\}$, (\ref{dsakj12837223gmis}) has been proved.

\section{Proof of the Approximation of $\phi(\lambda)$  for Small $\lambda$}\label{5464218penglua}
First, we propose a lemma below and this is the prerequisite for applying power series expansion on $\mathrm{exp}\left(-\lambda\Delta(k)\right)$  for small $\lambda$:
\begin{lemma}\label{das4541652lemma}
$\Delta(k)=\int_{\mathcal{C}}\left[1-F^k(s)-\left(1-F(s)\right)^K\right]\mathrm{d}\mathbf{p}_m$ is positive and bounded for $k\in\mathbb{N}^+$ and $k\leq K$.
\end{lemma}
\begin{IEEEproof}
Because $k\leq K$ and $F(s)$ is a CDF and thus satisfies $0<F(s)<1$, there exists a relation of the integrand in $\Delta(k)$:
\begin{equation}\label{yizhongxs}\small
1-F^k(s)-\left(1-F(s)\right)^K\geq 1-F^k(s)-\left(1-F(s)\right)^k.
\end{equation}

For $1-F^k(s)-\left(1-F(s)\right)^k$, we can employ mathematical induction as below to prove $1-F^k(s)-\left(1-F(s)\right)^k\geq0$:

For $k=1$, $1-F^1(s)-\left(1-F(s)\right)^1=0$ and the statement is true. Then for $k=n>1$, assuming $1-F^n(s)-\left(1-F(s)\right)^n\geq 0$ holds, we can have $1-F^n(s)\geq\left(1-F(s)\right)^n$. Because $0<F(s)<1$, we can further obtain
\begin{equation}\small
1-F^{n+1}(s)>1-F^n(s)\geq\left(1-F(s)\right)^{n}>\left(1-F(s)\right)^{n+1}.
\end{equation}
Therefore, the statement for $k=n+1$ is true and we thus prove the statement $1-F^k(s)-\left(1-F(s)\right)^k\geq 0$ for $k\in\mathbb{N}^+$. Due to (\ref{yizhongxs}) and $K\geq2$ (basic assumption of multicarrier systems), we thereby prove $1-F^k(s)-\left(1-F(s)\right)^K>0$ for $k\in\mathbb{N}^+$ and $k\leq K$. Meanwhile, because $F(s)$ will decrease exponentially with an increasing transmission distance, the area integral of the integrand $1-F^k(s)-\left(1-F(s)\right)^K$ over the region $\mathcal{C}$ is positive and bounded. As a result, the lemma is proved.
\end{IEEEproof}

By Lemma \ref{das4541652lemma}, $\mathrm{exp}\left(-\lambda\Delta(k)\right)$ is proved to be expandable for small $\lambda$ \cite{thomson2008elementary}. we can employ a power series expansion on $\mathrm{exp}\left(-\lambda\Delta(k)\right)$ for small $\lambda$ and obtain
\begin{equation}\small
\mathrm{exp}\left(-\lambda\Delta(k)\right)=1-\lambda\Delta(k)+\frac{\lambda^2\Delta^2(k)}{2}+O(\lambda^3).
\end{equation}

We can truncate $\mathrm{exp}\left(-\lambda\Delta(k)\right)$ by its second order term and substitute $\mathrm{exp}\left(-\lambda\Delta(k)\right)\approx 1-\lambda\Delta(k)+{\lambda^2\Delta^2(k)}/{2}$ into (\ref{45philamd65dsa451}), which yields
\begin{equation}\small
\begin{split}
\phi(\lambda)&\approx \sum_{k=1}^{K}\left[\binom{K}{k}(-1)^{k+1}\left(1-\lambda\Delta(k)+\frac{\lambda^2\Delta^2(k)}{2}\right)\right]\\
&=\sum_{k=1}^{K}\left[\binom{K}{k}(-1)^{k+1}\right]-\lambda\sum_{k=1}^{K}\left[\binom{K}{k}(-1)^{k+1}\Delta(k)\right]+\frac{\lambda^2}{2}\sum_{k=1}^{K}\left[\binom{K}{k}(-1)^{k+1}\Delta(k)^2\right].
\end{split}
\end{equation}

Furthermore, for $K\in\mathbb{N}^+$ and $K\geq 2$, by the binomial theorem, we can derive
\begin{equation}\label{dsa45d51542432hehesum}\small
\begin{split}
\sum_{k=1}^{K}\left[\binom{K}{k}(-1)^{k+1}\right]=1-\sum_{k=0}^{K}\binom{K}{k}(-1)^k=1-(1-1)^K=1.
\end{split}
\end{equation}
For the first order term, we can use the additivity property of integrals and swap the order of summation and integral by
\begin{equation}\label{54d65sa4561heszcs}\small
\begin{split}
&\sum_{k=1}^{K}\left[\binom{K}{k}(-1)^{k+1}\Delta(k)\right]=\int_{\mathcal{C}}\sum_{k=1}^{K}\left\lbrace\binom{K}{k}(-1)^{k+1}\left[1-F^k(s)-\left(1-F(s)\right)^K\right]\right\rbrace\mathrm{d}\mathbf{p}_m.
\end{split}
\end{equation}

Subsequently, the integrand can be determined by
\begin{equation}\small
\begin{split}
&\sum_{k=1}^{K}\left\lbrace\binom{K}{k}(-1)^{k+1}\left[1-F^k(s)-\left(1-F(s)\right)^K\right]\right\rbrace\\
&=\sum_{k=1}^{K}\left[\binom{K}{k}(-1)^{k+1}\right]-\sum_{k=1}^{K}\left[\binom{K}{k}(-1)^{k+1}F^{k}(s)\right]-(1-F(s))^K\sum_{k=1}^{K}\left[\binom{K}{k}(-1)^{k+1}\right]\\
&=1-[1-(1-F(s))^K]-(1-F(s))^K=0.
\end{split}
\end{equation}
As a result, the corresponding area integral of this zero integrand given in (\ref{54d65sa4561heszcs}) is also zero. Finally, the approximation of $\phi(\lambda)$ for small $\lambda$ given in (\ref{jinsidephilamd}) is proved.

\section{Proof of the Concavity of $\kappa(\tilde K,\lambda)$ in Terms of $\tilde K$}\label{2311profconcav}
According to the expressions of outage probability for bulk and per-subcarrier selections given in (\ref{phissbulk}) and (\ref{phissps}), it is obvious that $\Phi^{\Xi}(s)$ is a monotonically deceasing function of $\tilde K$. Therefore, $\forall~\beta\in[0,1]$, $\tilde K_1>0$ and $\tilde K_2>0$, we have
\begin{equation}\small
\Phi^{\Xi}(s)\vert_{K=(1-\beta)\tilde{K}_1+\beta\tilde{K}_2}\geq\Phi^{\Xi}(s)\vert_{K=(1-\beta)\tilde{K}_1}
\end{equation}
and
\begin{equation}\small
\Phi^{\Xi}(s)\vert_{K=(1-\beta)\tilde{K}_1+\beta\tilde{K}_2}\geq\Phi^{\Xi}(s)\vert_{K=\beta\tilde{K}_2}.
\end{equation}

As a result, it can be derived that
\begin{equation}\small
\begin{split}
\kappa((1-\beta)\tilde{K}_1+\beta\tilde{K}_2,\lambda)&\geq (1-\beta)\kappa((1-\beta)\tilde{K}_1,\lambda)+\beta\kappa(\beta\tilde{K}_2,\lambda).
\end{split}
\end{equation}
This proves the concavity of $\kappa(\tilde K,\lambda)$ in terms of $\tilde K$, according to the definition of a concave function \cite{arrow1961quasi}.

\bibliographystyle{IEEEtran}
\bibliography{bib}

\end{document}